\documentclass[conference]{IEEEtran}
\usepackage{cite}
\usepackage{amsmath,amssymb,amsfonts}
\usepackage{algorithm}
\usepackage{graphicx}
\usepackage{textcomp}
\usepackage{color, xcolor}
\usepackage{algpseudocode}
\usepackage{subfigure}
\usepackage{stfloats}
\usepackage{float}
\newtheorem{remark}{Remark}
\def\BibTeX{{\rm B\kern-.05em{\sc i\kern-.025em b}\kern-.08em
		T\kern-.1667em\lower.7ex\hbox{E}\kern-.125emX}}
\begin{document}

\title{Specific Beamforming for Multi-UAV Networks:\\ A Dual Identity-based ISAC Approach}

\author{
	\IEEEauthorblockN{Yanpeng~Cui\textsuperscript{*}, Qixun~Zhang\textsuperscript{*}, Zhiyong~Feng\textsuperscript{*}, Fan~Liu\textsuperscript{+}, Ce~Shi\textsuperscript{*}, Jinpo~Fan\textsuperscript{*}  and Ping~Zhang\textsuperscript{*}}
	\IEEEauthorblockA{\textsuperscript{*}Beijing University of Posts and Telecommunications, Beijing, P.R.China, 100876.\\
		\textsuperscript{+}Southern University of Science and Technology, Shenzhen, P.R.China, 518055.\\
		Email: \{cuiyanpeng94, zhangqixun, fengzy\}@bupt.edu.cn, liuf6@sustech.edu.cn, \{sc, fjp, pzhang\}@bupt.edu.cn}}
\maketitle

\begin{abstract}
Beam alignment is essential to compensate for the high path loss in the millimeter-wave (mmWave) Unmanned Aerial Vehicle (UAV) network. The integrated sensing and communication (ISAC) technology has been envisioned as a promising solution to enable efficient beam alignment in the dynamic UAV network. However, since the digital identity (D-ID) is not contained in the reflected echoes, the conventional ISAC solution has to either periodically feed back the D-ID to distinguish beams for multi-UAVs or suffer the beam errors induced by the separation of D-ID and physical identity (P-ID). This paper presents a novel dual identity association (DIA)-based ISAC approach, the first solution that enables specific, fast, and accurate beamforming towards multiple UAVs. In particular, the P-IDs extracted from echo signals are distinguished dynamically by calculating the feature similarity according to their prevalence, and thus the DIA is accurately achieved. We also present the extended Kalman filtering scheme to track and predict P-IDs, and the specific beam is thereby effectively aligned toward the intended UAVs in dynamic networks. Numerical results show that the proposed DIA-based ISAC solution significantly outperforms the conventional methods in association accuracy and communication performance.
\end{abstract}


\section{Introduction}
To provide wide coverage and on-demand services, the unmanned aerial vehicle (UAV) has been envisioned as a capable carrier for ubiquitous wireless intelligent communication \cite{UAV}. The millimeter-wave (mmWave) communication with its abundant spectrum resource has the potential to support the high-throughput and low-latency requirements of various UAV application scenarios. To compensate for the high path loss in mmWave-UAV communications, it is significant to generate narrow beams at the ground base station (BS) and align toward UAVs to achieve high array gains. However, it's not an easy task due to the highly dynamic mobility of UAVs.

The conventional schemes of aerial beam alignment rely on the UAV’s feedback. By transmitting a few pilots embedded in the downlink signal, the channel states can be estimated by UAVs and then fed back the optimal parameters to the BS. The feedback information includes the UAV's unique temporary identity (GUTI) and dynamic states (e.g., angle) required for the specific beam alignment. The former is more like a digital identity (D-ID) due to the pre-assigned and unique attributes, while the latter is more like a physical identity (P-ID) since it describes the time-varying mobility of UAVs. The feedback-based methods suffer from tedious feedback and the consequent delay; thus, the fleeting opportunities for beam alignment in highly dynamic environments may slip away. In view of this, research efforts toward predictive beamforming methods \cite{UAV_Learning_Beamforming1} are well underway.

In addition, several pioneering works have discussed the integration of sensing and communications (ISAC) for UAV networks \cite{UAV_ISAC1}  \cite{UAV_ISAC2}. The ISAC technique refers to a new information processing technology sharing information and software/hardware resources and realizes the coordination of sensing and communication functionalities. By exploiting the receivers' motion parameters from the reflected echoes, lower overhead and more accurate beam alignment have been realized \cite{UAV_ISAC3}. Nevertheless, a critical issue raised in the ISAC-based multi-UAV scenario is the separation of D-ID and P-ID since the D-ID is not contained in the echoes. This issue has been largely neglected and will induce severe consequences once D-ID and P-ID are wrongly associated: 1) the state estimation will be wrongly updated and 2) the information at each beam will be erroneously transmitted to the unintended UAV. Based on the Kullback-Leibler divergence, the distribution similarity of the estimated and predicted locations are compared in \cite{ID Association} to perform beam association. Since it only utilizes location information, periodical feedback is still inevitably required to avoid association error when the trajectory crosses radially. By exploiting all the available features, a Euclidean distance-based scheme is developed in \cite{Com_Served_By_Sen} to associate the beam to the corresponding receiver. However, the numerical results are shown based on perfect beam association, which can not be guaranteed in some scenarios such as formation flight. Despite the significance of dual identity association (DIA) in multi-UAV networks, it has not been well investigated.

In this paper, we present a novel DIA-based ISAC approach, which is the first solution that enables the specific beams to be accurately and swiftly aligned toward the intended UAVs. In particular, the reflected echo signal is exploited for both beam tracking and prediction, which removes the overhead of downlink pilot and uplink feedback and reduces the beam tracking delay. Besides, it also brings significant matched-filtering gain and preserves most of the angular information to reduce estimation errors, resulting in an accurate beam alignment. To distinguish UAVs' P-IDs in dynamic environments, the measured features are assigned with dynamic weights according to the prevalence to calculate similarity. We also resort to our recent work, the vampire bat optimizer (VBO), to associate the P-ID pairs measured and predicted by minimizing the similarity difference. We also develop the extended Kalman filtering (EKF) for tracking and predicting the P-IDs, and the efficient and predictive beamforming design is realized. Simulation results show that our solution is significantly superior to the conventional methods in both identity association accuracy and communication performance.

\textit{Notations}: Unless otherwise specified, matrices and vectors are denoted by bold uppercase and italic bold lowercase letters, respectively. Besides, the term $\mathbb{C}^{A\times B}$ denotes a complex space of $A\times B$ dimension, and the subscripts indicate the indexes of the time slot and number of UAVs, e.g., $\mathbf{p}_{k,n}$ is the location vector of the $k$th UAV at the $n$th epoch.

\section{System Model}
\subsection{Signal and Communication Model}
\begin{figure}[h]
	\centering
		\includegraphics[width=0.8\linewidth]{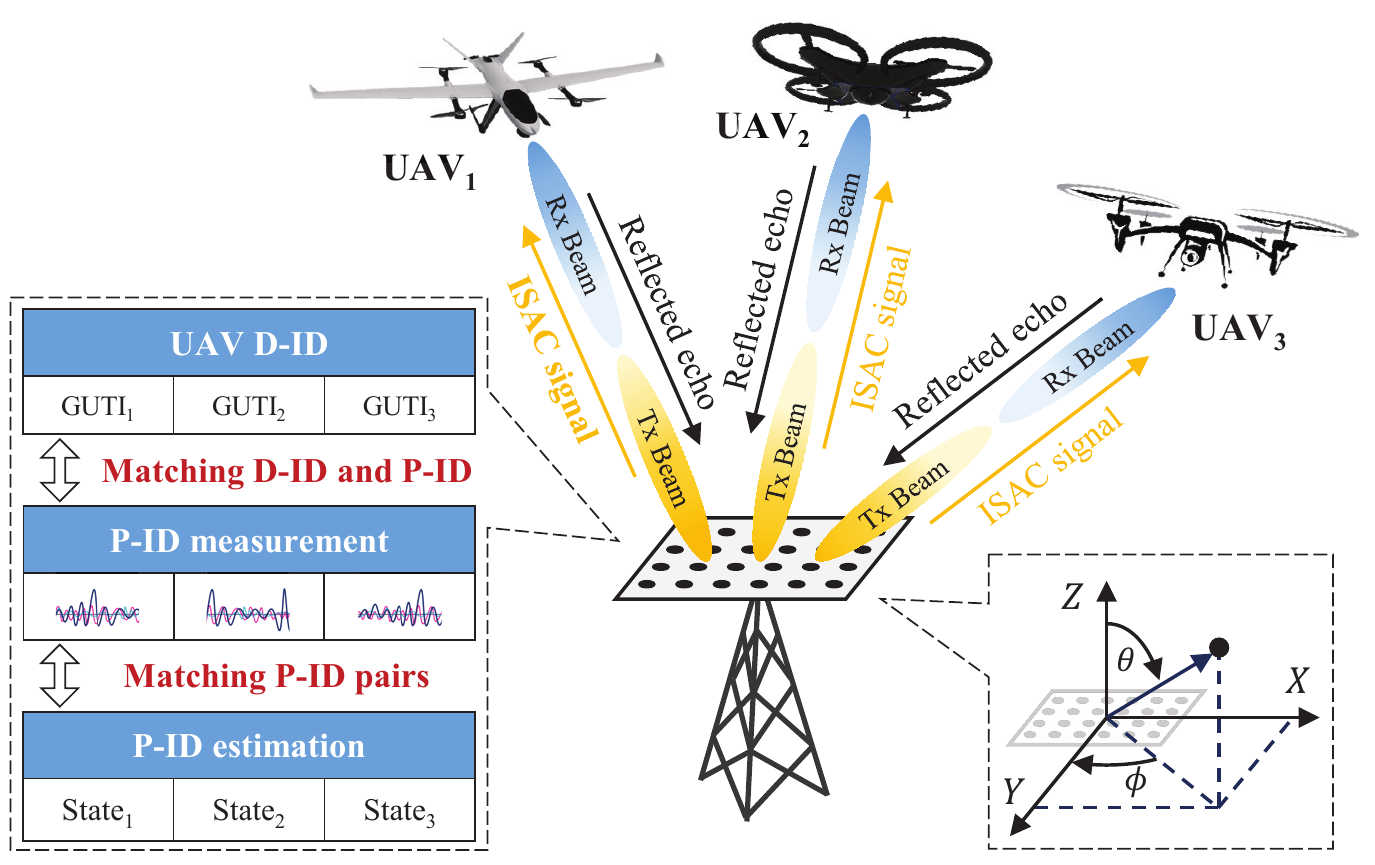}
	\caption{The typical scenario of the mmWave multi-UAV networks.}
	\label{Scenario}
\end{figure}

As shown in Fig. \ref{Scenario}, we consider a cellular-connected multi-UAV network consisting of $K$ UAVs. The ground BS is equipped with a mmWave massive MIMO (mMIMO) uniform planar array (UPA) consisting of $N_t$ transmit antennas and $N_{r_b}$ receive antennas. By exploiting full-duplex radio techniques on transmit and receive antennas\cite{Full_Duplex}, the signal echoes for sensing can be received while maintaining uninterrupted downlink communications concurrently\footnote{We assume that the self-interference issue at BS can be addressed by the separate transmit and receive UPA, with strong radio frequency isolation combined with highly directive elevation beamforming \cite{Self}. To avoid alleviating from the focus of this paper, it will be designated as our future work and not be discussed in this paper.}. The locations of the BS and the $k$th UAV at the $n$th slot are denoted as $\mathbf{p}_b$ and $\mathbf{p}_{k,n}=[{\rm p}_{k,n}(1),{\rm p}_{k,n}(2),{\rm p}_{k,n}(3)]^T$, where ${\rm p}_{k,n}(i),i=1,2,3$ represent the $k$th UAV's locations on axis $x$, $y$ and $z$, respectively. Moreover, the UPA is also assumed to be deployed at the bottom of UAVs.

By denoting the multi-beam ISAC signals toward $K$ UAVs as $\textbf{s}_n(t)=[s_{1,n}(t),…,s_{K,n}(t)]^T\in\mathbb{C}^{K\times1}$, the transmitted
signals are given by
$\tilde{\textbf{s}}_n(t)=\mathbf{F}_n\textbf{s}_n(t)\in\mathbb{C}^{N_t\times1}$, where $N_t=N_{tx}\times N_{ty}$ denotes the number of transmit antennas of BS, $\mathbf{f}_{k,n}=\mathbf{a}(\phi_{k,n},\theta_{k,n})$ is the $k$th column of the beamforming matrix $\mathbf{F}_n$. The signal $\tilde{\textbf{s}}_{k,n}(t)=\mathbf{f}_{k,n}s_{k,n}(t)$ is received by the $k$th UAV via a receive beamformer $\mathbf{w}_{k,n}$, yielding
\begin{equation}\label{Received_Signal} 
	\begin{aligned}
	r_{k,n}(t)=\ &A_{k,n}\mathbf{w}_{k,n}^H\mathbf{b}(\phi_{k,n},\!\theta_{k,n})\mathbf{a}^H(\phi_{k,n},\!\theta_{k,n})\tilde{\textbf{s}}_{k,n}(t)\\&+\mathbf{z}_r(t)\textbf{}
\end{aligned},
\end{equation}
where $\mathbf{z}_r(t)$ is the zero-mean Gaussian noise (hereinafter referred to as noise) with variance $\sigma_r^2$. According to \cite{LoS1}, the communication channels between UAVs and BS are mainly dominated by the line of sight (LoS) links. Thus we have $A_{k,n}=\sqrt{N_tN_rp_{k,n}}\tilde{\alpha}d_{k,n}^{-1}e^{j\frac{2\pi f_c}{c}d_{k,n}}$, where $p_{k,n}$ denotes the transmit power, $N_r$ denotes the number of receive antennas of UAV. $\tilde{\alpha}d_{k,n}^{-1}$ denotes the path-loss of the LoS channel with $\tilde{\alpha}$ being the power gain at unit reference distance. $2\pi f_cd_{k,n}/c$ denotes the phase of the channel, where $d_{k,n}$, $f_c$ and $c$ are the distance of the $k$th UAV relative to the BS, carrier frequency and the speed of light, respectively. $\mathbf{a}(\phi,\theta)$ represents the transmit steering vectors of the BS's UPA, which is given by
\begin{equation}\label{Steering_Vector}
	\begin{aligned}
	\mathbf{a}(\phi,\theta)=\sqrt{\frac{1}{N_t}}\big[&1,\ldots,e^{j\pi {\rm sin}\theta[(n_a-1){\rm cos}\phi+(n_b-1){\rm sin}\phi]},\\ &\ldots,e^{j\pi {\rm sin}\theta[(N_a-1){\rm cos}\phi+(N_b-1){\rm sin}\phi]}\big]^T
\end{aligned},
\end{equation}
where we assume the UPA has half-wavelength antenna spacing. $\mathbf{b}(\phi,\theta)$ is the receive steering vector of the UAV's UPA and is similarly defined as (\ref{Steering_Vector}) with $N_{r_u}$ antennas. 

Assuming that the ISAC signal has a unit power, the receive signal-to-noise (SNR) is expressed as $\Gamma_{k,n}=p_{k,n}|A_{k,n}\mathbf{w}_{k,n}^H\mathbf{b}_{k,n}(\phi,\theta)\mathbf{a}_{k,n}^H(\phi,\theta)\mathbf{f}_{k,n}|^2/\sigma_a^2$,
and the average achievable rate is $R_n=\frac{1}{K}\sum_{k=1}^{K}{{\rm log}_2\left(1+\Gamma_{k,n}\right)}$.

\subsection{Radar Measurement Model}
According to the mMIMO theory \cite{mMIMO_Theory}, we have the following mathematical lemma $|\mathbf{a}_{k,n}^H(\phi,\theta)\mathbf{a}_{k,n}(\phi^{\prime},\theta^{\prime})|\!\to\!0,\forall \phi\!\neq\!\phi^{\prime},\theta\!\neq\!\theta^{\prime}, N_t\!\!\to\!\!\infty$. In other words, the beams formulated by the mMIMO UPA of BS will be adequately narrow, thus the steering vectors are nearly orthogonal to each other. Consequently, no inter-beam interference exists, namely the echoes from various UAVs will not interfere with each other and thus they can be individually processed by the BS. For the $k$th UAV, the reflected echo received at the BS has the following expression as
\begin{equation}\label{Echo} 
	\begin{aligned}
	\mathbf{c}_{k,n}(t)&=B_{k,n}\mathbf{u}(\phi_{k,n},\!\theta_{k,n})\mathbf{a}^H\!(\phi_{k,n},\!\theta_{k,n})\tilde{\textbf{s}}_{k,n}(t-\tau_{k,n})\\&\ \ \ +\mathbf{z}_c(t)
\end{aligned}, 
\end{equation}
where \!$B_{k,n}\!\!=\!\!\sqrt{N_tN_{r_b}p_{k,n}}\beta_{k,n}e^{j2\pi\mu_{k,\!n}\!t}$, with $N_{r_b}$,\! $\beta_{k,n}$,\! $\mu_{k,n}$ and $\tau_{k,n}$ being the number of receive antennas, the reflection coefficient, the Doppler frequency and the time-delay, respectively. $\mathbf{u}(\phi_{k,n},\!\theta_{k,n})$ denotes the steering vector of BS's receive UPA, which is similarly defined as (\ref{Steering_Vector}) with $N_{r_b}$ antennas. $\mathbf{z}_c(t)$ represents the noise with variance $\sigma^2$.

The transmit SNR of (\ref{Echo}) is defined as $p_{k,n}/\sigma_c^2$. The $\mu_{k,n}$ and $\tau_{k,n}$ can be estimated by matched-filtering (\ref{Echo}) with a Doppler-shifted delayed version of $s_{k,n}(t)$ \cite{Variance}. Therefore, $\mathbf{p}_{k,n}$ and $\mathbf{v}_{k,n}$ are measured by
\begin{equation}\label{Range_and_Velocity}
	\left\{
	\begin{aligned}
		&\hat{\tau}_{k,n}=2||\mathbf{p}_{k,n}-\mathbf{p}_b||/c+z_\tau \\
		&\hat{\mu}_{k,n}=2\mathbf{v}_{k,n}^T(\mathbf{p}_{k,n}-\mathbf{p}_b) f_c/(c||\mathbf{p}_{k,n}-\mathbf{p}_b||)+z_f
	\end{aligned}
	\right.,
\end{equation}
where $\mathbf{v}_{k,n}=[{\rm v}_{k,n}(1),{\rm v}_{k,n}(2),{\rm v}_{k,n}(3)]^T$ denotes the velocity of the $k$th UAV, $z_\tau$ and $z_f$ denote the noise with variance $\sigma_1^2$ and $\sigma_2^2$, respectively. $||\cdot||$ is the norm of a vector. By compensating (\ref{Echo}) with the estimation and normalizing the results by $p_{k,n}$ and the matched-filtering gain $G$, we obtain
\begin{equation}\label{Echo2} 
	{\tilde{\mathbf{c}}}_{k,n}\!=\!\sqrt{N_tN_{r_b}}\beta_{k,n}\mathbf{u}(\phi_{k,n},\!\theta_{k,n})\mathbf{a}^H\!(\phi_{k,n},\!\theta_{k,n})\mathbf{f}_{k,n}+\mathbf{\tilde{z}}_c.
\end{equation}
Furthermore, the estimation of the reflection coefficient is realized via ${\hat{\beta}}_{k,n}=\xi/\hat{\tau}_{k,n}c$, which depends on the radar cross-section $\xi$ of UAV. The likelihood function $p({\tilde{\mathbf{c}}}_{k,n}|\phi_{k,n},\theta_{k,n})$ is defined based on (\ref{Echo2}), and both $\hat{\phi}_{k,n}$ and $\hat{\theta}_{k,n}$ can be finally obtained via the maximum likelihood estimator
\begin{equation}\label{MLE} 
	\{\hat{\phi}_{k,n},\hat{\theta}_{k,n}\}=\mathop{arg\max}\limits_{\phi_{k,n}\in\Phi,\theta_{k,n}\in\Theta}{p\left({\tilde{\mathbf{c}}}_{k,n}|\phi_{k,n},\theta_{k,n}\right)},
\end{equation}
where $\Phi$ and $\Theta$ denote the set containing all possible values of $\hat{\phi}_{k,n}$ and $\hat{\theta}_{k,n}$, respectively. The variance of the measuring noise of $\phi_{k,n}$ and $\theta_{k,n}$ are denoted as $\sigma_3^2$ and $\sigma_4^2$, respectively.

According to \cite{Variance}, variance $\sigma_i^2, i=1\sim4$ are inversely proportional to the receive SNR of (\ref{Echo}), and we thus assume that $\sigma_i^2=a_i^2\sigma^2/(GN_tN_{r_b}|\beta_{k,n}|^2|\mathbf{a}_{k,n}^H(\phi,\theta)\mathbf{f}_{k,n}|^2p_{k,n}),i=1,2$, and $\sigma_i^2=a_i^2\sigma^2/(Gp_{k,n}),i=3,4$, where $a_i, \forall i$ depend on the system configuration, etc.

\section{Predictive and Accurate Beamforming by the DIA-based ISAC Approach}

\begin{figure}
	\centering
	\includegraphics[width=0.9\linewidth]{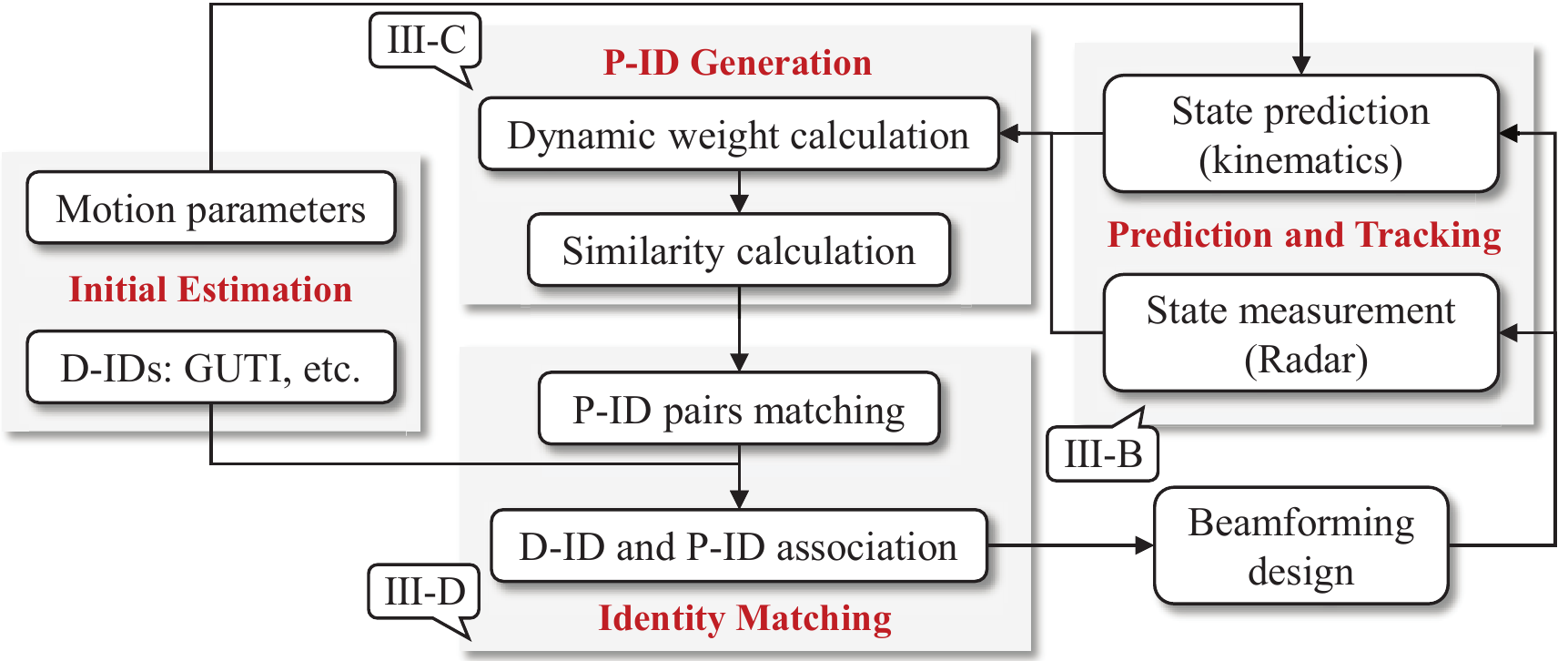}
	\caption{The framework of the proposed DIA solution.}
	\label{Framework}
\end{figure}
As shown in Fig. \ref{Framework}, the framework of our proposed DIA approach is mainly composed of four modules: i) initial estimation, ii) beam prediction and tracking, iii) P-ID generation and iv) identity matching.
\subsection{Initial Estimation}
Our solution is initialized by estimating UAVs' motion parameters at BS, which can be easily realized by conventional beam training. In this stage, the association between D-IDs and P-IDs is perfectly obtained since they are fed back from the same signal from a specific UAV. This initial association is the precondition for the subsequent association since the conventional feedback is replaced by echo sensing, from which we can only obtain P-ID rather than D-ID.

\begin{remark}
	It should be noted that the D-ID is not limited to the above-mentioned GUTI, any unique label, e.g., IP/MAC address, can be utilized to represent D-ID. In addition, any distinguishable feature is included in its connotation of P-ID. In addition to the distance, velocity and angles introduced in Sec. \uppercase\expandafter{\romannumeral2}, other motion parameters can also be exploited for P-ID generation, e.g., the unique micro-Doppler frequency caused by blade numbers and rotor speeds can be extracted and exploited to distinguish UAVs. We choose not to discuss them in this article to avoid alleviating from the focus of DIA to parameter estimation.
\end{remark}

\subsection{Beam Prediction and Tracking}
After the initial access process for beam alignment, the state prediction and tracking will continue for a long period of time $T$, which is discretized into several small time slots $\Delta T$. The signal transmitted by the BS is received by the UAV's UPA and also reflected by the fuselage. The reflected echos are exploited to measure the motion parameters by (\ref{Range_and_Velocity})$\sim$(\ref{MLE}) at the $n-1$th epoch and perform one- and two-step prediction by using the kinematic equations of the UAVs. More details will be presented in Sec. \uppercase\expandafter{\romannumeral4}-A. By using the one-step predictions at the $n$th epoch, the BS formulates transmit beamformer $\mathbf{f}_{k,n}$ based on $\mathbf{\hat{\theta}}_{k,n|n-1}$ and $\mathbf{\hat{\phi}}_{k,n|n-1}$. Each of the beams contains the information of the two-step predictions $\mathbf{\hat{\theta}}_{k,n+1|n-1}$ and $\mathbf{\hat{\phi}}_{k,n+1|n-1}$ to avoid the one-step information becoming outdated at the $n+1$th epoch. As a result, UAVs will formulate receive beamformer $\mathbf{w}_{k,n+1}$ based on the two-step prediction. The beams between BS and UAVs will be aligned once the prediction is accurately obtained. 

\subsection{P-ID Generation}
After obtaining the measurement and prediction of the states, the BS further generates the distinguishable P-ID to make a unique physical identification for each UAV. Specifically, it denotes $\textbf{PF}_n=\left\{\textbf{\textbf{pf}}_{k,n}, k=1,...,K\right\}$ as the observed feature vector of $K$ UAVs, where $\textbf{\textbf{pf}}_{k,n}(m)$ denotes the $m$th physical feature ($m=1,...,M$) of the $k$th UAV. The P-IDs are established by collecting all the observable and distinguishable physical features and formulating dynamic weights based on their prevalence. More details will be presented in Sec. \uppercase\expandafter{\romannumeral4}-B.
\subsection{Identity Matching}
Identity matching has the following two connotations: i) associating the P-IDs obtained at different epochs, which is called P-ID pair matching, and ii) associating D-ID and P-ID at the same epoch to ensure that the specific beam can be aligned toward the intended UAV. Recall that the initial association result of D-ID and P-ID has been obtained in the initial estimation stage. After that, the matching accuracy of D-ID and P-ID can be ensured in the long term once the P-ID pairs of any two adjacent time slots are iteratively associated accurately. More details will be discussed in Sec. \uppercase\expandafter{\romannumeral4}-C.
\begin{figure}
	\centering
	\includegraphics[width=1\linewidth]{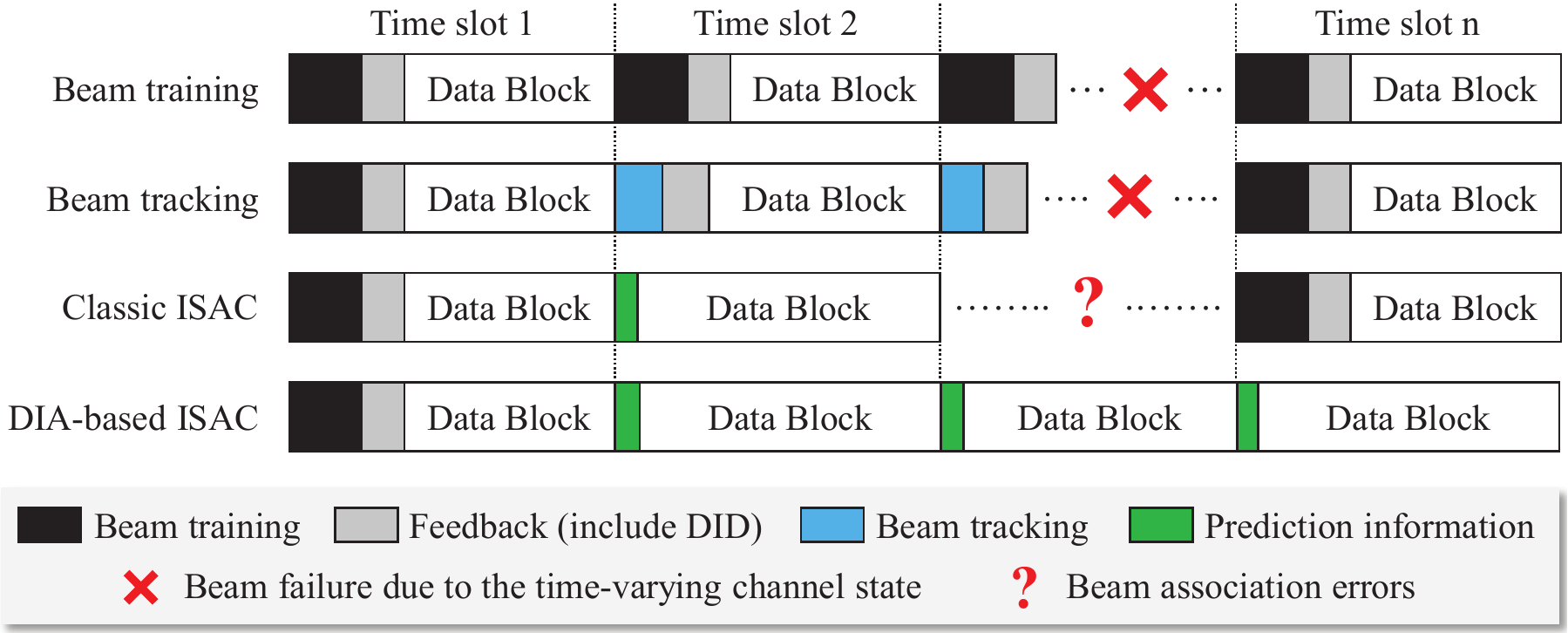}
	\caption{The transmission block structure of the proposed method.}
	\label{Frame_Structure}
\end{figure}

For clarification, we compare the frame structure of the conventional beam training/tracking methods, the classic ISAC, and our proposed DIA-based ISAC solution in Fig. \ref{Frame_Structure}. For the beam training/tracking methods, there are no beam association errors since the feedback brings the D-ID information. However, the downlink pilots and uplink feedback are both indispensable, resulting in large overhead and time consumption. In addition, the angles information learned at the latest epoch may be outdated, resulting in that beams may not align owing to the time-varying channel states. The classic ISAC solution removes the pilot overhead and the frequent feedback. Whereas, unless periodically introducing the D-ID feedback, which is contrary to the original intention of ISAC design, it will still suffer from beam association error in multi-UAV networks since the D-ID information is not contained in the reflected echo. Our proposed DIA-based ISAC solution inherited the advantages of the classic ISAC scheme, including the low overhead and high efficiency. Furthermore, after obtaining the D-ID information in the initial stage, the BS will subsequently track the UAVs and associate their D-IDs with the P-IDs extracted from the echoes. As a result, the dedicated resources reserved for D-ID feedback can all be saved for transmitting useful data without any beam mismatching.

\section{Predictive and Accurate DIA}
In what follows, we will detail the technique we designed for the proposed DIA-based ISAC approach and the discussion about computational complexity.

\subsection{Beam Prediction and Tracking}

In this subsection, a Kalman filtering scheme is proposed for beam prediction and tracking. Due to the non-linearity in the measurement function (\ref{Range_and_Velocity})$\sim$(\ref{MLE}), the linear Kalman filtering can not be utilized directly. We thus develop an EKF method that performs linearization for nonlinear measurement. 

The state variables and measured vectors are denoted as $\boldsymbol{x}_{k,n}=[\,\mathbf{p}_{k,n},\mathbf{v}_{k,n}]^T$ and $\boldsymbol{y}_{k,n}=[\tau_{k,n},\mu_{k,n},\tilde{\mathbf{c}}_{k,n}^T]^T$. Following the standard assumption in \cite{UAV_Learning_Beamforming1}, the motions of UAV are regarded to keep constant within $\Delta T$, the models of state evolution and measurement can be given by $\boldsymbol{x}_n=\textbf{G}\boldsymbol{x}_{n-1}+\boldsymbol{u}_{n-1}$ and $\boldsymbol{y}_n=\textbf{H}(\boldsymbol{x}_{n})+\boldsymbol{z}_{n}$,
where $\textbf{G}=\left[\textbf{I}_{3\times3},\Delta T\cdot\textbf{I}_{3\times3};\textbf{0}_{3\times3},\textbf{I}_{3\times3}\right]$, and $\textbf{H}(\cdot)$ is defined as (\ref{Range_and_Velocity})$\sim$(\ref{MLE}). $\boldsymbol{u}$ and $\boldsymbol{z}$ are noises with covariance matrices as $\textbf{Q}_s={\rm diag}(\sigma_{p(i)}^2,\sigma_{v(i)}^2), i=1,2,3$ and $\textbf{Q}_m={\rm diag}(\sigma_{1}^2,\sigma_{2}^2,[\sigma_{3}^2,\sigma_{4}^2]\textbf{1}_{N_{rb}}^T)$, $\textbf{1}_{N_{rb}}^T$ is a size-$N_{rb}$ all-one vector.

In order to linearize the measurement models, let's denote $\boldsymbol{\eta}(\theta)=\sqrt{N_tN_{r_b}}\beta\mathbf{u}(\phi,\!\theta)\mathbf{a}^H\!(\phi,\!\theta)\mathbf{a}(\hat{\phi},\!\hat{\theta})$, where $\hat\theta$ and $\hat\phi$ are predictions for $\theta$ and $\phi$. The Jacobian matrix for $\textbf{H}(\boldsymbol{x})$ is
\begin{equation}\label{jacobian}
\dfrac{\partial \textbf{H}}{\partial \boldsymbol{x}}=\left[\begin{matrix}m(1)\!\!\!&m(2)\!\!\!&m(3)\!\!\!&0\!\!\!&0\!\!\!&0\\b(1)\!\!\!&b(2)\!\!\!&b(3)\!\!\!&\tilde{m}(1)\!\!\!&\tilde{m}(2)\!\!\!&\tilde{m}(3)\\q(1)\!\!\!&q(2)\!\!\!&q(3)\!\!\!&0\!\!\!&0\!\!\!&0\end{matrix}\right],
\end{equation}
where $b(i)=\frac{2f_c[{\rm v}(i)-{\rm p}(i)\mathbf{p}^T\mathbf{v}]}{c\sqrt{\mathbf{p}^T\mathbf{p}}}$, $q(i)=\frac{\partial\eta}{\partial\theta}\frac{{\rm p}(i){\rm p}(3)}{\mathbf{p}^T\mathbf{p}\sqrt{\mathbf{p}^T\mathbf{p}-{\rm p}^2(3)}}$, $m(i)=\frac{2{\rm p}(i)}{c\sqrt{\mathbf{p}^T\mathbf{p}}}$, $\tilde{m}(i)=f_cm(i)$, $i=1,2,3$. The partial derivative of $\boldsymbol{\eta}$ with respect to $\theta$ is given by
\begin{equation}\label{partial_eta}
	\frac{\partial\boldsymbol{\eta}}{\partial\theta}=\frac{\beta}{\sqrt{N_t}}\left[\begin{matrix}\sum\limits_{i=1}^{N_{tx}}\!\sum\limits_{j=1}^{N_{ty}}\dfrac{\psi_{\hat{\theta},\hat{\phi}}(i,j,1,1)\chi_{\theta,\phi}(i,j,1,1)}{\psi_{\theta,\phi}(i,j,1,1)}\\\sum\limits_{i=1}^{N_{tx}}\!\sum\limits_{j=1}^{N_{ty}}\dfrac{\psi_{\hat{\theta},\hat{\phi}}(i,j,1,1)\chi_{\theta,\phi}(i,j,2,2)}{\psi_{\theta,\phi}(i,j,2,2)}\\...\\\sum\limits_{i=1}^{N_{tx}}\!\sum\limits_{j=1}^{N_{ty}}\dfrac{\psi_{\hat{\theta},\hat{\phi}}(i,j,1,1)\chi_{\theta,\phi}(i,j,N_{ry},N_{rx})}{\psi_{\theta,\phi}(i,j,N_{ry},N_{rx})}\\\end{matrix}\right],
\end{equation}
where $\psi_{\alpha,\beta}(i,j,\zeta_1,\zeta_2)=e^{j\pi {\rm sin}\alpha\left[\left(j-\zeta_1\right){\rm cos}\beta+\left(i-\zeta_2\right){\rm sin}\beta\right]}$ and $\chi_{\alpha,\beta}(i,j,\zeta_1,\zeta_2)=\frac{\partial {\rm ln}(\psi_{\alpha,\beta}(i,j,\zeta_1,\zeta_2))}{\partial\theta}$.

We are now ready to present the EKF procedure, and the state prediction and tracking steps are summarized as follows.
\textit{1) State prediction:}
$\hat{\boldsymbol{x}}_{n|n-1}=\textbf{G}\boldsymbol{x}_{n-1}$, $\hat{\boldsymbol{x}}_{n+1|n-1}=\textbf{G}\boldsymbol{x}_{n|n-1}. $
\textit{2) Linearization:}
${\rm \textbf{H}}_n=\left.\frac{\partial \textbf{H}}{\partial \boldsymbol{x}}\right|_{\boldsymbol{x}=\boldsymbol{\hat{x}}_{n|n-1}}.$
\textit{3) Prediction of the mean squared error (MSE) matrix:}
${\rm \textbf{M}}_{n|n-1}={\rm \textbf{G}}_{n-1}{\rm \textbf{M}}_{n-1}{\rm \textbf{G}}_{n-1}^H+{\rm \textbf{Q}}_s.$
\textit{4) Calculation of the Kalman gain:}
${\rm \textbf{K}}_n={\rm \textbf{M}}_{n|n-1}{\rm \textbf{H}}_{n}^H({\rm \textbf{Q}}_{m}+{\rm \textbf{H}}_{n}{\rm \textbf{M}}_{n|n-1}{\rm \textbf{H}}_{n}^H)^{-1}.$
\textit{5) State tracking:}
$\hat{\boldsymbol{x}}_{n}=\hat{\boldsymbol{x}}_{n|n-1}+{\rm \textbf{K}}_n(\boldsymbol{y}_{n}-{\rm \textbf{H}}(\hat{\boldsymbol{x}}_{n|n-1})).$
\textit{6) MSE matrix update:}
${\rm \textbf{M}}_{n}=({\rm \textbf{I}}-{\rm \textbf{K}}_{n}{\rm \textbf{H}}_{n}){\rm \textbf{M}}_{n|n-1}.$

By performing prediction and tracking iteratively, the BS can simultaneously sense and
communicate with $K$ UAVs according to the optimal angles.

\begin{remark}
To provide the best angles for the beamformer, the predicted location $\mathbf{\hat{p}}_{k,n}$ is utilized to calculate the predicted angles via ${\rm p}_{k,n}(1)=d_{k,n}{\rm sin}\theta_{k,n}{\rm sin}\phi_{k,n}$, ${\rm p}_{k,n}(2)=d_{k,n}{\rm sin}\theta_{k,n}{\rm cos}\phi_{k,n}$ and ${\rm p}_{k,n}(3)=d_{k,n}{\rm cos}\phi_{k,n}$, which may have conversion errors and thus lead to biased estimation. As for the unbiased converted measurements method, readers can refer to our recent work \cite{UAV_ISAC2} \cite{Dual_Identity} and we won't reiterate it here.
\end{remark}

\subsection{P-ID Generation}
This subsection proposed a dynamic P-ID generation method to provide the subsequent P-ID association stage with a reliable similarity metric. Given the prediction and estimation of UAV's states, we want to calculate $S_n(i,j)$, namely the similarity of the measurements $\boldsymbol{y}_{n}$ and their estimations $\hat{\boldsymbol{y}}_{n|n-1}$. Nevertheless, it is not trivial since their distinguishability depends on the dynamic environment. For instance, locations are more distinctive in a low-density network, while velocity has low importance in distinguishing when UAVs are in formation. This motivates us to dynamically compute the weight for features based on their prevalence.

We exploit a cosine function $C\left(\textbf{\textbf{pf}}_n(a),\textbf{\textbf{pf}}_n(b)\right)=\textbf{\textbf{pf}}_n^T(a)\textbf{\textbf{pf}}_n(b)/\left\|\textbf{\textbf{pf}}_n(a)\right\|\left\|\textbf{\textbf{pf}}_n(b)\right\|$ to describe the similarity of two features, namely how likely they belong to the same UAV. 
We denote $\textbf{\textbf{pf}}_n(m)=\{\textbf{\textbf{pf}}_{k,n}(m), k=1,...,K\}$ as the vector of all measurements of the $m$th physical feature on $K$ UAVs. 
The weight of the $m$th feature is assigned as $w_{n}(m)=\frac{1}{K}\sum\nolimits_{k=1}^{K}P_{k,n}(m)$,
 where $P_{k,n}(m)=\sum\nolimits_{j\neq k}C(\textbf{\textbf{pf}}_{k,n}(m), \textbf{\textbf{pf}}_{j,n}(m))\prod\nolimits_{q\neq j,q\neq k}(1-C(\textbf{\textbf{pf}}_{k,n}(m), \textbf{\textbf{pf}}_{q,n}(m)))$ is the distinguishability of $\textbf{\textbf{pf}}_{k,n}(m)$, namely the probability that $\textbf{\textbf{pf}}_{k,n}(m)$ is different from other features in $\textbf{\textbf{pf}}_n(m)$. Then the similarity of $\textbf{\textbf{pf}}_{i,n}$ and $\textbf{\textbf{pf}}_{j,n}$ is defined as the harmonic mean of individuals 
\begin{equation}\label{Similarity} 
	S_n(i,j)=\left\{\sum\limits_{m=1}^{M}\frac{w_n^{\prime}(m)}{C\left(\textbf{\textbf{pf}}_{i,n}(m), \textbf{\textbf{pf}}_{j,n}(m)\right)}\right\}^{-1}, 
\end{equation}
where $w_n^{\prime}(m)$ denotes the normalized weight, and $M$ is the number of observable features extracted from measurement. 

The above methods dynamically weigh various features via their prevalence. Besides, if two UAVs have high dissimilarities in most features, their similarity will be low despite the large weights of other features, and the outliers could also be mitigated. As a result, the similarity matrix that is more convenient for P-ID pairs association can be established, and the details will be introduced in the next subsection. 

\subsection{Identity Matching}
This subsection details the necessity and procedure for identity matching. Note that the original association of UAV's D-ID and P-ID has been obtained at the initial access stage. In the subsequent process, the P-ID acquisition is achieved by echo signal processing at BS instead of communication feedback from UAVs. However, since the D-ID is not contained in the echos, the BS should have the capability to distinguish multiple UAVs by associating the subsequently measured P-ID with the D-ID obtained at the initial access stage to realize the following two aims: i) correctly aligning the specific beam toward the intended UAV and ii) correctly updating the state estimation introduced in Sec \uppercase\expandafter{\romannumeral4}-A. This implies that each UAV's P-ID information must be accurately correlated in any two adjacent time slots. To tackle this issue, we propose an efficient P-ID pair matching approach.

\begin{figure}
	\vspace{0.15in}
	\centering
	\includegraphics[width=0.9\linewidth]{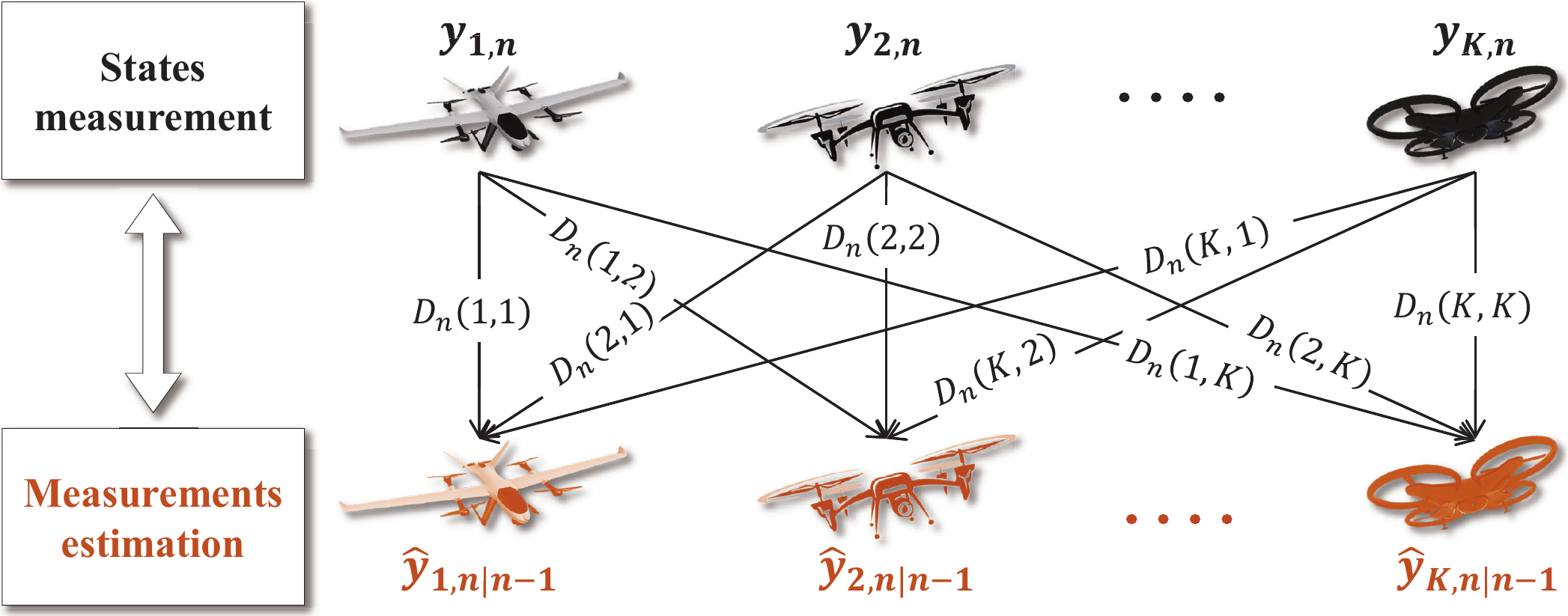}
	\caption{Bipartite graph model for matching P-ID pairs}
	\label{Matching_Model}
\end{figure}

Recall that the BS detects $K$ UAVs and formulate measurements $\boldsymbol{y}_{k,n}, k=1,...,K$, we calculate all the measurement estimation of state predictions by $\hat{\boldsymbol{y}}_{j,n|n-1}=\textbf{H}(\hat{\boldsymbol{x}}_{j,n|n-1}),j=1,...,K$. The measurement $\boldsymbol{y}_{i,n}$ will not be ``far from" $\hat{\boldsymbol{y}}_{j,n|n-1}$, so we calculated their difference, namely the reciprocal of their similarity by $D_n(i,j)=S_n(i,j)^{-1}$, and then establish the bipartite graph matching model
shown in Fig. \ref{Matching_Model}, where the edge's weight is defined as the matching cost $D_n(i,j),\forall i,j$. 

The optimization target of the P-ID pairs matching problem is to minimize the overall and the individual differences, that is, $\min(f_1+f_2)$, where the sub-targets of the overall cost and the individual cost are denoted as $f_1=\frac{1}{K}\sum\nolimits_{i=1}^{K}\sum\nolimits_{j=1}^{K}A_n(i,j)D_n(i,j)$ and $f_2=\frac{1}{K}\sqrt{\sum\nolimits_{i=1}^{K}\sum\nolimits_{j=1}^{K}\left(A_n(i,j)D_n(i,j)-f_1\right)^2}$, respectively. The above optimization target is constrained to $\sum\nolimits_{i=1}^{K}A_n(i,j)=1$ and $\sum\nolimits_{j=1}^{K}A_n(i,j)=1$, $i,j=1,...,K$, namely any $\boldsymbol{y}_{i,n}$ or $\hat{\boldsymbol{y}}_{j,n|n-1}$ can only be used to match once. $\textbf{\textit{A}}_n$ is the assignment matrix, where $A_n(i,j)=1$ if $\boldsymbol{y}_{i,n}$ is assigned to $\hat{\boldsymbol{y}}_{j,n|n-1}$, otherwise $A_n(i,j)=0$. 

The above-mentioned matching problem of optimal association of P-ID pairs can be solved by VBO, which is an effective optimizer we recently proposed. The specific steps will not be introduced here due to page limitations, and readers can refer to our recent work \cite{Vampire Bat Optimizer} for more details.

By doing so, based on the matching results of P-ID pairs and the association between D-ID and P-ID after initial access, the D-IDs are successfully matched with P-IDs during the subsequent beam tracking process. As a result, the BS can transmit the correct signal to the intended UAVs at each beam, which achieves the aforementioned aim i). In addition, the BS can update the state prediction $\hat{\boldsymbol{x}}_{n|n-1}$ based on the correct measurement, which achieves the aim ii). 

\subsection{Computational Complexity}
In the beam prediction and tracking stage, the EKF requires the execution matrix inversion, having a cubic complexity order of the state vector dimension $V$, i.e., $O(V^3)$.
The computational complexity of P-ID generation lies in the dynamic similarity calculation, which is $O(MK^2)$. In the identity matching stage, the VBO has the order of complexity $O( K^2{\rm log}(KC))$ to minimize the globe cost and equalize the local cost, where $C$ is the maximum absolute value of competition. Considering that the $V$ and $M$ are significantly smaller than ${\rm log}(KC)$ in the large-scale UAV network, the order of computational complexity of the proposed DIA solution can be thus regarded as that of the VBO.

\section{Numerical Results}

In this section, we present the numerical results to validate the effectiveness of the proposed DIA solution. Let's consider a network with 10$\sim$20 UAVs moving freely in 3-D spaces. Their initial positions are randomly generated on a hemispherical surface with a radius of 100 m, and the BS is located at the center of the sphere. All UAVs select a horizontal direction towards the BS and randomly deviates within 10 degrees and a vertical direction that is randomly distributed within $\pm$10 degrees. The lower bound of speed is 8 m/s, and the upper bound is between 10 m/s and 30 m/s. The BS is operating in $f_c=28$ GHz \cite{28GHz}. The duration for each time slot is $\Delta T=0.02s$. We set $G=10$, $\sigma=\sigma_r=1$, $\tilde{\alpha}=1$, $\sigma_{p(i)}=0.02$ m, $\sigma_{v(i)}=0.2$ m/s, $a_1=6.7\times10^{-7}$, $a_2=2\times10^4$, $a_3 = a_4 = 1$, respectively. We run 40 Monte Carlo trials to evaluate the performance, and the average results are discussed below.
\begin{figure}
	\centering
	\includegraphics[width=0.8\linewidth]{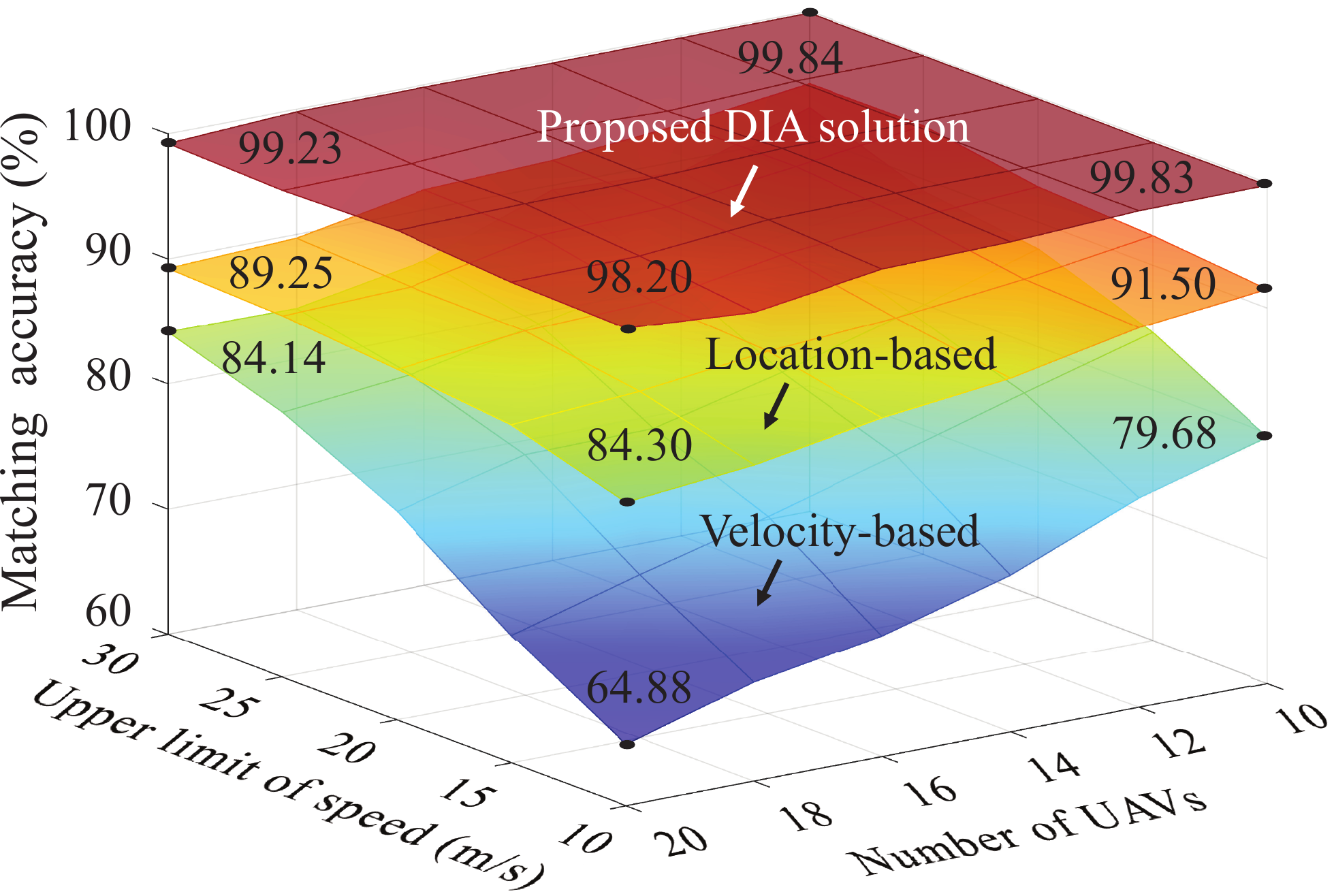}
	\caption{Beam matching performance for the proposed DIA solution and the velocity- and location-based schemes.}
	\label{Matching_Accuracy}
\end{figure}

\subsection{Performance of Association Accuracy}
In Fig. \ref{Matching_Accuracy}, we compare the matching accuracy (the proportion of beams that correctly match the intended UAVs) for different solutions. The velocity-based solution which only utilizes the speeds and heading directions to distinguish UAVs has the worst performance since they might be repeatedly similar during the long-term flight. In addition, with the reduction of the upper bound of speed and the increase of the UAV number, it suffers more matching errors since there will be more UAVs with similar velocities. The location-based solution also suffers from matching errors since UAVs may occasionally approach each other due to the free flight. Such cases are rarer than those with similar velocity, thus the performance of the location-based one is slightly better than the velocity-based one. Our DIA solution can almost attain the matching accuracy of 100\% since the P-ID of UAV is generated according to the feature prevalence. That is, UAVs' locations play a more significant role in the matching cost matrix when their velocities are similar, and vice versa. Averagely, it outperforms the location- and velocity-based solutions by 18.67\% and 9.24\%, respectively.

Since there are still cases where the velocity and location are simultaneously indistinguishable, our DIA solution does not achieve the matching accuracy of 100\%. However, an average gap of about 0.61\% from perfect matching is still acceptable owing to the fact that we only utilize two features to formulate P-ID, and the experimental results are sufficient to prove the effectiveness of the dynamic weight. In the future, the full utilization of all available features (e.g., the micro-Doppler characteristics) is expected to approach the perfect matching performance since it is extremely rare that all features are simultaneously indistinguishable.
\begin{figure}
	\centering
	\includegraphics[width=0.9\linewidth]{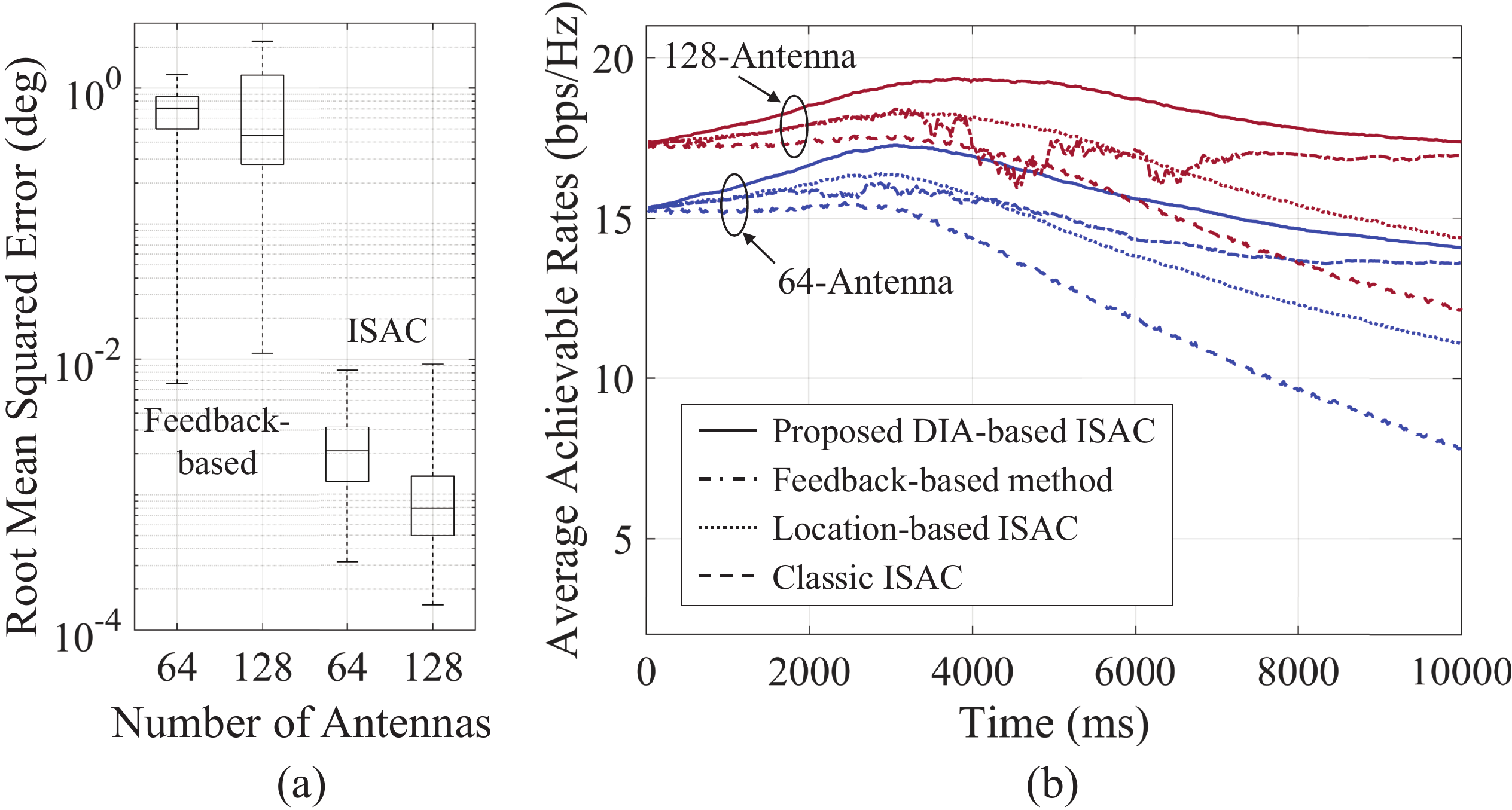}
	\caption{Performance comparison of angle estimation and achievable rate. The upper limit of speed is 20 m/s, $K=10$, $N_t=N_{r_{u}}=N_{r_{b}}=64\ {\rm or}\ 128$. }
	\label{Achievable_Rate}
\end{figure}

\subsection{Performance of Communication}
In Fig. \ref{Achievable_Rate}, we show the tracking error of angle and the average achievable rates for the feedback-based solution, the classic ISAC solution \cite{UAV_ISAC3}, the location-based ISAC solution \cite{ID Association} and our DIA-based ISAC solution. It should be noted that the beam association errors of the last three solutions are not considered in Fig. \ref{Achievable_Rate}(a), so they are classified into one category, namely ISAC. In addition, in the statistics of Fig. \ref{Achievable_Rate}(b), the achievable rate of a link will be zero if the P-ID of a UAV is not correctly associated with its D-ID, i.e., a beam is wrongly associated with an unintended UAV. 

Most UAVs will fly over the top of the BS in the time region 3000$\sim$5000 ms, all solutions achieve the maximum rates since the communication distance is the shortest. Nevertheless, note that the angles change too fast during this period, and the rate of the feedback-based solution drops owing to the following reasons: i) there is only one single pilot being exploited for tracking and ii) it requires a receive beamformer to combine the pilot signal and inevitably leads to the loss of the detailed angle information. As shown in Fig. \ref{Achievable_Rate}(a), the angle estimations have large errors and become even worse in the 128-antenna scenario given the narrower beam and higher misalignment probability. The ISAC-based solution utilizes the whole echo signal block for sensing, and the matched filter gain is 10 times of that in the feedback-based one. In addition, it does not perform receive beamforming for the echos and thus most of the angular information can be preserved, resulting in the best achievable rates.

In addition, the proposed DIA-based ISAC solution maintains the highest rates since it has almost no matching errors. The feedback-based scheme notifies the BS of the UAV's D-ID in real-time, so there is no beam failure problem caused by mismatch and its rates follow our solution closely. For the classic ISAC solution, beam failure issue frequently occurs since D-ID is not contained in the echoes. According to statistics, 6 of 10 beams of UAVs are wrongly associated on average, so the rates will gradually decrease to about 40\% of our solution. The location-based ISAC solution only utilizes location information to associate UAVs and beams. It is gradually worse than the feedback-based one since it suffers more than one matching error in each trial as shown in Fig. \ref{Matching_Accuracy}, resulting in a 20\% degradation of average rates. 

\subsection{Performance of Time Consumption}
\begin{figure}[t]
	\centering
	\subfigure[Transmitter]{\includegraphics[width=0.45\columnwidth]{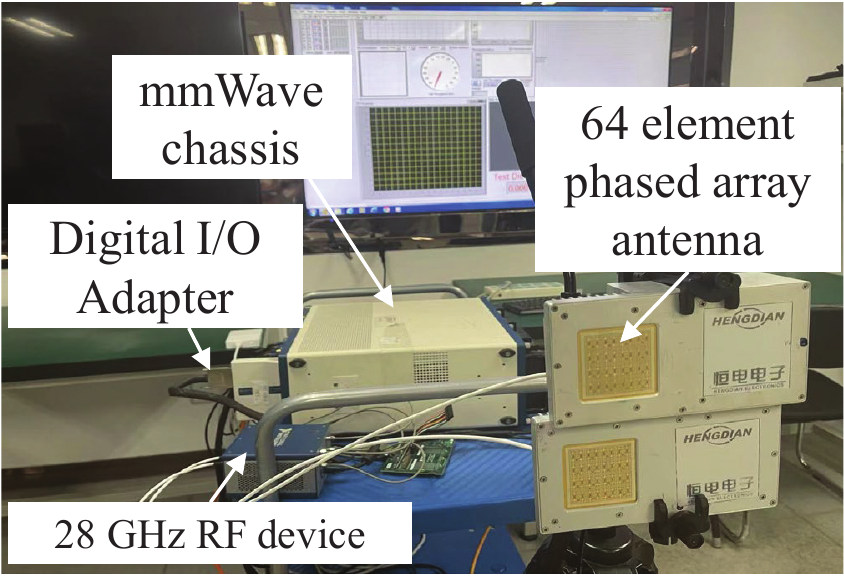}}
	\subfigure[Receiver (moving target)]{\includegraphics[width=0.45\columnwidth]{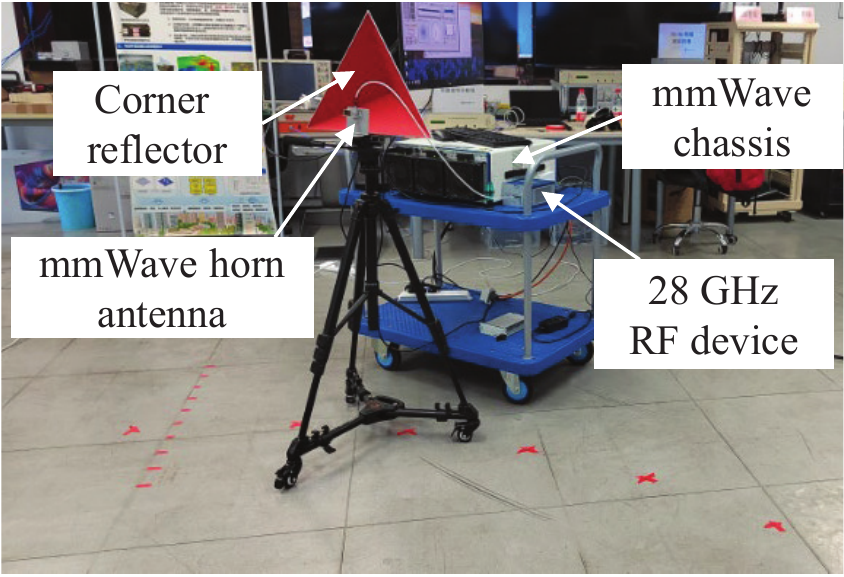}}
	\caption{Real-world experiment environment: the hardware testbed for time-division ISAC system.}\label{Testbed}
\end{figure}
\begin{figure}
	\centering
	\includegraphics[width=0.9\linewidth]{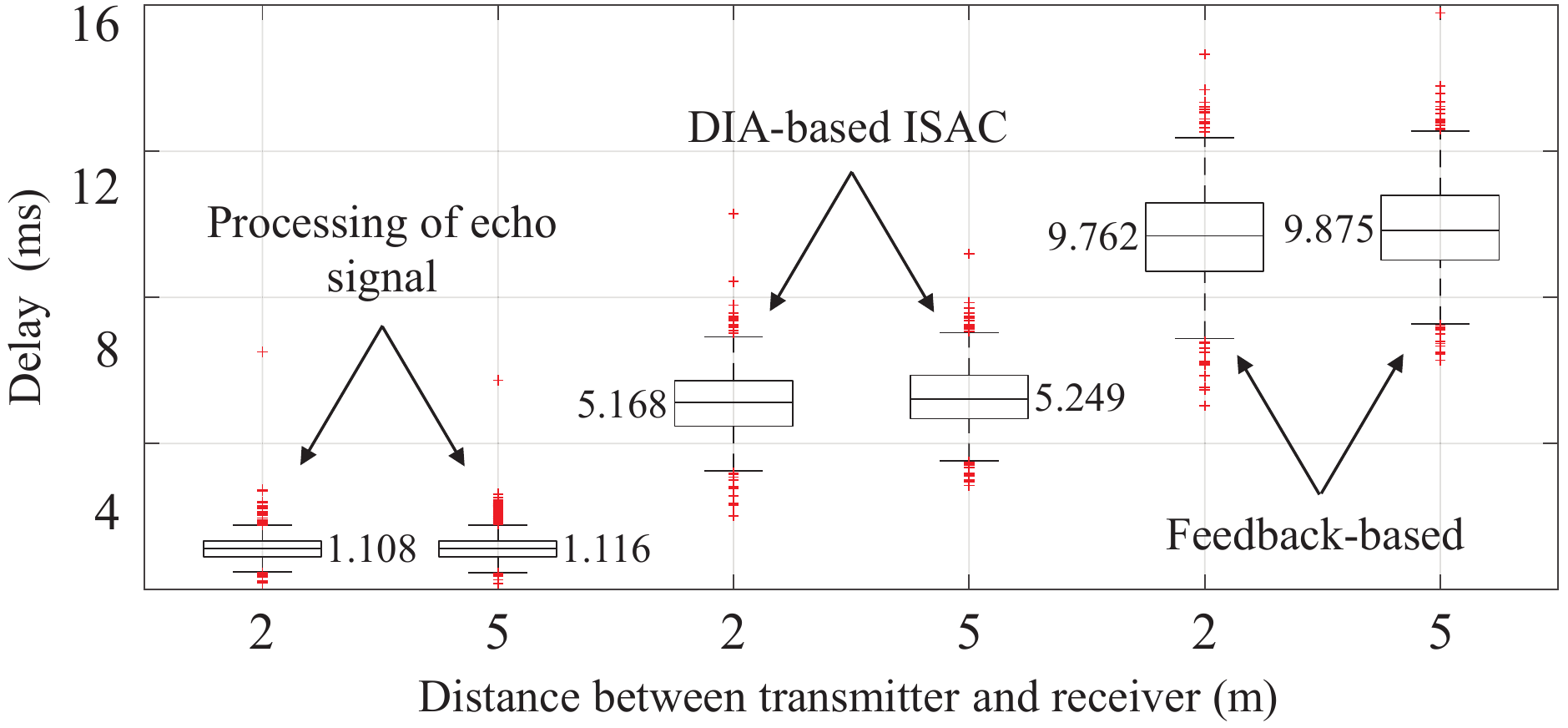}
	\caption{Performance comparison of time consumption of echo signal processing, the proposed DIA-based ISAC solution, and the feedback-based solution.}
	\label{Time_Consumption}
\end{figure}
In view of the improvement of matching accuracy and achievable rate, other performances may be sacrificed. Therefore, we evaluate one of them, i.e., the time consumption, with the hardware testbed shown in Fig. \ref{Testbed}, which has been established in our recent work \cite{testbed}. The specific hardware design will not be introduced here due to page limitations, and readers can refer to \cite{testbed} for more details. The target is moving with a speed of 1 m/s, and the distance varies from 2 m to 5 m. We perform 4096 points Inverse Fast Fourier Transform (IFFT) radar signal processing on the reflected echo. As shown in Fig. \ref{Time_Consumption}, compared with the feedback-based one, the DIA-based ISAC solution reduces the total delay by 4.594 ms and 4.626 ms in 2 m and 5 m, respectively. The unique echo processing delay only takes about 1.1 ms. In addition, as analyzed in Sec. \uppercase\expandafter{\romannumeral4}-D, the complexity of the proposed algorithm other than VBO can be ignored. The matching result of 10 nodes with two features can be obtained within 0.5 ms on average based on VBO, and this is the only additional time consumption when compared with the classic ISAC method. Although the experiment is performed indoors rather than in an aerial communication scene, it is sufficient to make conclusions that our proposed DIA-based ISAC approach outperforms the feedback-based solution in terms of time consumption and is not much inferior to the classic ISAC.

\section{Conclusion}
In this paper, the proposed DIA-based ISAC approach distinguishes features according to their prevalence, tracks the beam by the EKF method, and matches dual identity accurately in dynamic UAV networks. As a result, without the tedious feedback, the specific narrow beams can be swiftly and correctly aligned towards the intended receiver in multi-UAV networks. Simulation results demonstrated that the proposed approach significantly outperforms the conventional methods in association accuracy and communication performance. The systematical evaluation of beam alignment in actual scenarios will be designated as our future work.
\section*{Acknowledgment}

This work was partly supported by Major Research Projects of the National Natural Science Foundation of China (92267202), the National Key Research and Development Project (2020YFA0711303), and the BUPT Excellent Ph.D. Students Foundation (CX2022208).

\vspace{12pt}


\begin{thebibliography}{1}
\bibitem{UAV}
Y. Cui, Q. Zhang et al., ``Topology-Aware Resilient Routing Protocol for FANETs: An Adaptive Q-Learning Approach," \emph{IEEE Internet Things J.}, vol. 9, no. 19, pp. 18632-18649, Oct. 2022.
	
\bibitem{UAV_Learning_Beamforming1}
C. Liu, W. Yuan, et al., ``Location-Aware Predictive Beamforming for UAV Communications: A Deep Learning Approach," \emph{IEEE Wireless Commun. Lett.}, vol. 10, no. 3, pp. 668-672, Mar. 2021.
	
	
\bibitem{UAV_ISAC1}
Z. Lyu, G. Zhu and J. Xu, ``Joint Trajectory and Beamforming Design for UAV-Enabled Integrated Sensing and Communication," in \emph{Proc. IEEE International Conference on Communications (ICC)}, Seoul, Korea, May 2022, pp. 1593-1598.

\bibitem{UAV_ISAC2}
Y. Cui, Q. Zhang, Z. Feng et al., ``Toward Trusted and Swift UAV Communication: ISAC-Enabled Dual Identity Mapping," \emph{IEEE Wireless Communications}, 2023, Accepted.


\bibitem{UAV_ISAC3}
Z. Wei, F. Liu, D. W. Kwan Ng and R. Schober, ``Safeguarding UAV Networks through Integrated Sensing, Jamming, and Communications," in \emph{Proc. IEEE International Conference on Acoustics, Speech and Signal Processing (ICASSP)}, Singapore, May 2022, pp. 8737-8741.

\bibitem{ID Association}
Z. Wang, K. Han, et al., ``Multi-Vehicle Tracking and ID Association Based on Integrated Sensing and Communication Signaling," \emph{IEEE Wireless Commun. Lett.}, vol. 11, no. 9, pp. 1960-1964, Sept. 2022.

\bibitem{Com_Served_By_Sen}
F. Liu, W. Yuan, et al., ``Radar-Assisted Predictive Beamforming for Vehicular Links: Communication Served by Sensing," \emph{IEEE Trans. Wireless Commun.}, vol. 19, no. 11, pp. 7704-7719, Nov. 2020. 

\bibitem{Full_Duplex}
C. B. Barneto, S. D. Liyanaarachchi, et al., ``Full duplex radio/radar technology: The enabler for advanced joint communication and sensing," \emph{IEEE Wireless Commun.}, vol. 28, no. 1, pp. 82–88, Feb. 2021.

\bibitem{Self}
W. Yuan, F. Liu, C. Masouros, J. Yuan, D. W. K. Ng and N. González-Prelcic, ``Bayesian Predictive Beamforming for Vehicular Networks: A Low-Overhead Joint Radar-Communication Approach," \emph{IEEE Trans. Wireless Commun.}, vol. 20, no. 3, pp. 1442-1456, March 2021.

\bibitem{LoS1}
Z. Xiao et al., "A Survey on Millimeter-Wave Beamforming Enabled UAV Communications and Networking," in \textit{IEEE Commun. Surveys Tuts}., vol. 24, no. 1, pp. 557-610, 1st Quart., 2022.


\bibitem{mMIMO_Theory}
H. Q. Ngo, \textit{Massive MIMO: Fundamentals and System Designs}, vol. 1642. Linköping, Sweden: Linköping Univ. Electronic Press, 2015.
 
\bibitem{Variance}
S. M. Kay, \textit{Fundamentals of Statistical Signal Processing: Estimation Theory}. vol. 1. Englewood Cliffs, NJ, USA: Prentice-Hall, 1998.

\bibitem{Dual_Identity}
Y. Cui, Q. Zhang, Z. Feng et al., ``Dual Identities Enabled Low-Latency Visual Networking for UAV Emergency Communication," in \emph{Proc. IEEE Global Communications Conference (GLOBECOM)}, Brazil, Dec. 2022, pp. 474-479.

\bibitem{Vampire Bat Optimizer}
X. Zhao, Y. Cui et al., ``Energy-Efficient Coverage Enhancement Strategy for 3-D Wireless Sensor Networks Based on a Vampire Bat Optimizer,'' \emph{IEEE Internet Things J.}, vol. 7, no. 1, pp. 325-338, Jan. 2020.

\bibitem{28GHz}
Y. Wang, T. Phelps, et al., ``28 GHz 5G-Based Phased-Arrays for UAV Detection and Automotive Traffic-Monitoring Radars," in \emph{Proc. IEEE/MTT-S International Microwave Symposium (IMS)}, Philadelphia, PA, USA, Aug. 2018, pp. 895-898.

\bibitem{testbed}
Q. Zhang et al., ``Time-Division ISAC Enabled Connected Automated Vehicles Cooperation Algorithm Design and Performance Evaluation,'' \emph{IEEE J. Sel. Areas Commun.}, vol. 40, no.7, pp. 2206-2218, July 2022.


\end{thebibliography}
\end{document}